\documentclass[aps,prb,showpacs,twocolumn,superscriptaddress,floatfix]{revtex4}

\usepackage{graphicx}
\usepackage{dcolumn}
\usepackage{bm}
\usepackage{amsmath}
\usepackage{amssymb}

\begin{document}

\renewcommand{\vec}[1]{\mbox{\boldmath $#1$}}


\title{Neutron scattering study of role of partial disorder-type spin fluctuations in conductivity of frustrated conductor Mn$_3$Pt}


\author{K. Tomiyasu}
\email[Electronic address: ]{tomiyasu@m.tohoku.ac.jp}
\affiliation{Department of Physics, Tohoku University,
Aoba, Sendai 980-8578, Japan}
\author{H. Yasui}
\affiliation{Power and Industrial Systems R{\&}D Center, TOSHIBA Corporation,
Toshiba-cho, Fuchu-shi, Tokyo 183-8511, Japan}
\author{Y. Yamaguchi}
\affiliation{IMR, Tohoku University,
Aoba, Sendai 980-8577, Japan}


\date{\today}

\begin{abstract}
The spin-frustrated conductor Mn$_3$Pt exhibits a characteristic magnetic structure called partial disorder in which some spin sites can form magnetic order through the generation of non-ordered sites that locally relieve the frustration. Here we report the results of a single-crystal inelastic neutron scattering study of this compound. The measured momentum $\vec{Q}$ correlations of diffusive magnetic scattering reveal that the paramagnetic phase exhibits short-range spin fluctuations with the same type of partial disorder. Its relation to conductivity is also discussed.
\end{abstract}


\maketitle

\section{Introduction}
%
On a triangular lattice and a tetrahedral lattice, only some classical-spin pairs can be arranged antiferromagnetically.~\cite{Wannier_1950,Anderson_1956} This leads to a situation called geometrical frustration, which suppresses magnetic ordering and leads to intriguing phenomena such as liquid-like short-range spin fluctuations in spin ices and spin molecules~\cite{Bramwell_2001,Lee_2002} and the formation of complex magnetic structures with multiferroics in insulators.~\cite{Tomiyasu_2004,Yamasaki_2006} However, recently, molecular spin/spin-orbit excitations were also discovered even in a magnetically ordered phase, which forms when frustration is relieved, demonstrating that spin dynamics is indispensable to frustration.~\cite{Tomiyasu_2008,Tomiyasu_2011}

Frustrated conductors can exhibit novel transport properties. Examples of such properties include heavy fermion behavior in the spinel LiV$_2$O$_4$ and in C15 Laves phase compound Y(Sc)Mn$_2$ \cite{Matsushita_2005,Shiga_1993}; superconductivity in the pyrochlore oxide Cd$_2$Re$_2$O$_7$ \cite{Hanawa_2001}; and metal-insulator (MI) transitions in the pyrochlores Cd$_2$Os$_2$O$_7$, Hg$_2$Ru$_2$O$_7$, and $R_{2}$Ir$_2$O$_7$ ($R =$ Nd, Sm, Eu, Gd, Tb, Dy, or Ho) \cite{Sleight_1974,Yamamoto_2007,Matsuhira_2011}. All these systems consist of corner-sharing tetrahedral lattices of magnetic atoms. Thus, one of the challenging issues in current condensed matter physics is to determine the relation between the transport properties and frustration.

\begin{figure}[htbp]
\begin{center}
\includegraphics[width=0.85\linewidth, keepaspectratio]{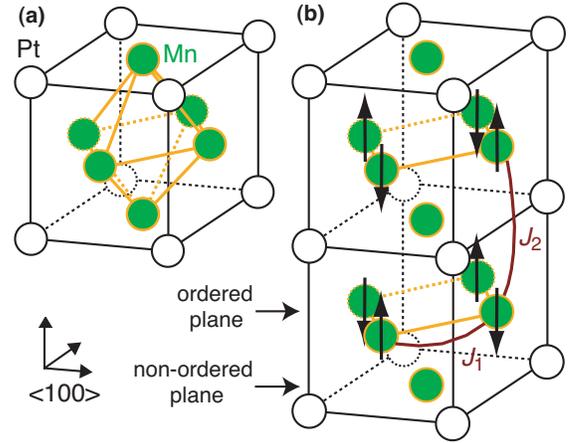}
\end{center}
\caption{\label{fig:str} (Color online) Structures of unit cell in Mn$_3$Pt. (a) Crystal structure. Mn atoms form an octahedron. (b) Magnetic structure in F-phase. The first- and second-neighbor magnetic interactions $J_1$ and $J_2$ are also defined. }
\end{figure}

The Cu$_3$Au-type intermetallic compound Mn$_3$Pt is also a conductor with spin frustration. As shown in Fig.~\ref{fig:str}(a), nonmagnetic Pt atoms and magnetic Mn atoms occupy corner and face centers, respectively. Here, the Mn atoms form a rare type of octahedral lattice, neither triangular nor tetrahedral, providing a basis for the occurrence of frustration.
With decreasing temperature from the high-temperature paramagnetic phase, this compound transitions into an antiferromagnetic F-phase at $T_{\rm N}=475$ K with a tiny tetragonal contraction of $c/a=0.9985$, and then enters into an antiferromagnetic D-phase at $T_{\rm t}\simeq400$ K with a large volume contraction of 2.25{\%}.~\cite{Kren_1968,Yasui_1992,Ikeda_2003}
Interestingly, the F-phase exhibits partial disorder in its magnetic structure, which is described by a propagation vector $\vec{q}_{\rm F}=(1,0,1/2)$ in a cubic notation, as shown in Fig.~\ref{fig:str}(b).~\cite{Kren_1968} The antiferromagnetically ordered square-lattice plane with 2.2$\mu_{\rm B}$ and the non-ordered plane are alternately stacked along the $\langle100\rangle$ direction, where $\mu_{\rm B}$ is the Bohr magneton. Partial disorder is a representative frustration effect:~\cite{Mekata_1977} non-ordered sites appear so as to relieve frustration, allowing the other sites to form spin order.
In addition, in both the paramagnetic phase and F-phase, very diffuse and quasielastic magnetic scattering was found roughly along the first Brillouin zone boundary by single-crystal neutron scattering, indicating the existence of characteristic short-range spin fluctuations arising from frustration.~\cite{Ikeda_2003}

Although none of the aforementioned dramatic transport properties have been reported in Mn$_3$Pt, its electrical conductivity exhibits rather interesting behavior: conductivity drops at $T_{\rm N}$ with decreasing temperature.~\cite{Yasui_1987} Normally, a paramagnetic phase should exhibit lower conductivity than a magnetically ordered phase owing to spin-flop scattering. Thus, frustration is possibly responsible for the anomalous conductivity in some form. However, to resolve this question, information on the spatial correlations of short-range spin fluctuations, a key feature of magnetic frustration, will be necessary.

In this study, we performed single-crystal inelastic neutron scattering experiments on Mn$_3$Pt. Momentum $\vec{Q}$ correlations of the short-range spin fluctuations were investigated with much higher statistics and over a much wider $\vec{Q}$ space than in the previous study.~\cite{Ikeda_2003} On the basis of the results, we studied the spatial spin correlations and developed a plausible explanation for the frustration effect on conductivity.

\section{Experiments}
Single-crystal inelastic neutron scattering experiments were performed on the triple-axis spectrometer TOPAN installed at the JRR-3 reactor, JAEA, Tokai, Japan. The final energy of the neutrons was fixed at $E_{\rm f}=14.7$ meV with a horizontal collimation sequence of blank-100$^{\prime}$-100$^{\prime}$-blank. A sapphire filter and a pyrolytic graphite filter efficiently eliminated fast neutrons and higher-order contamination, respectively.
As samples, single-crystal specimens of Mn$_3$Pt were grown by the Bridgeman method.
Each crystal was about 10 mm in diameter and 25 mm tall. Two co-aligned single crystals were fixed by thin aluminum plates and enclosed in an aluminum container, which was placed under the cold head of a $^{4}$He closed-cycle refrigerator with a high-temperature option.

\section{Results}
\begin{figure}[htbp]
\begin{center}
\includegraphics[width=0.95\linewidth, keepaspectratio]{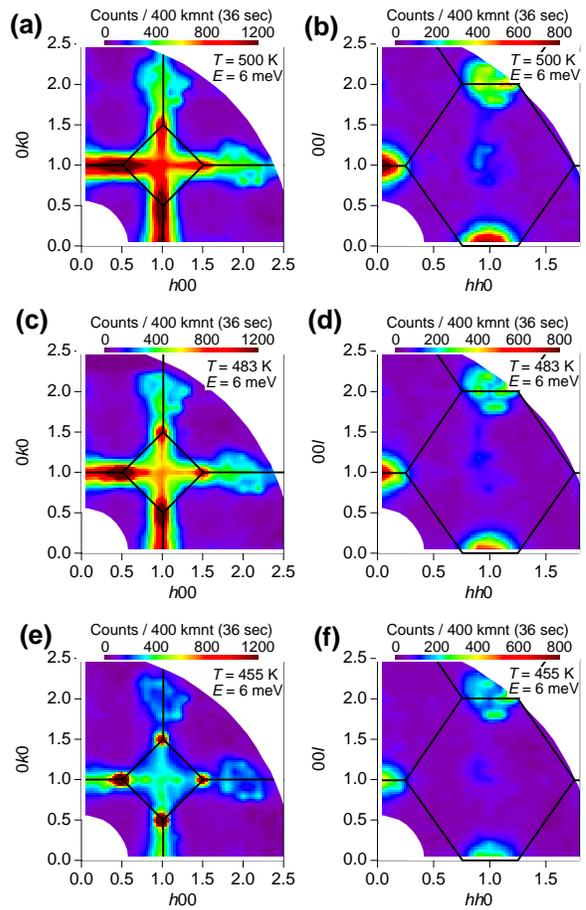}
\end{center}
\caption{\label{fig:CEmaps} (Color online) Color images of single-crystal inelastic neutron scattering data. These were acquired in the $hk$0 and $hhl$ zones in a constant-energy scan mode at 6 meV. (a)(b), (c)(d), and (e)(f) were acquired in the paramagnetic phase, at just above $T_{\rm N}$ in the paramagnetic phase, and in the F-phase, respectively. }
\end{figure}

Figures~\ref{fig:CEmaps}(a) to \ref{fig:CEmaps}(f) show the measured $\vec{Q}$ correlations of spin fluctuations in the paramagnetic phase and F-phase.
At 500 K (in the paramagnetic phase), as shown in Fig.~\ref{fig:CEmaps}(a), diffuse scattering is observed not only along the first Brillouin zone boundary, but at points distributed along a cross-like pattern in the $hk$0 zone.
The intensity is relatively strong around 1 1/2 0, 1/2 1 0, 1 3/2 0, and 3/2 1 0, which coincide with the magnetic Bragg reflection points described by $\vec{q}_{\rm F}$ in the F-phase. This means that the short-range correlations of the F-phase magnetic structure dynamically survive.
In the $hhl$ zone, as shown in Fig.~\ref{fig:CEmaps}(b), the spot-like diffuse scattering is observed around 110 and 001, corresponding to parts of the cross-like pattern. This indicates that the cross-like pattern does not spread perpendicularly to the $hk$0 zone, but consists of two rods described by $h$10 and 1$k$0. Such rod-type scattering is known as a sign of two-dimensional (2D) correlations such as in high-$T_{\rm c}$ cuprates.

At 483 K, just above $T_{\rm N}$, as shown in Figs.~\ref{fig:CEmaps}(c) and ~\ref{fig:CEmaps}(d), the overall intensity decreases and concentrates toward the $\vec{q}_{\rm F}$ points in the $hk$0 zone.
At 455 K (in the F-phase), as shown in Figs.~\ref{fig:CEmaps}(e) and \ref{fig:CEmaps}(f), the cross-like pattern in the $hk$0 zone and the 110 and 001 spots in the $hhl$ zone are quite weak. The concentrated $\vec{q}_{\rm F}$ spots are more condensed, most likely due to a spin wave that stems from the magnetic structure.

The ring-like patterns are also observed around 120 and 210 in the $hk$0 zone, as shown in Figs.~\ref{fig:CEmaps}(a), \ref{fig:CEmaps}(c), and \ref{fig:CEmaps}(e), and around 112 in the $hhl$ zone, as shown in Figs.~\ref{fig:CEmaps}(b), \ref{fig:CEmaps}(d), and \ref{fig:CEmaps}(f). However, no similar patterns are observed around the equivalent points 100, 010, and 110, which are associated with lower $|\vec{Q}|$. Therefore, the rings are identified as being phononic, not magnetic,  in nature since the former and latter neutron scattering intensities decrease and increase with decreasing $|\vec{Q}|$ because of a $|\vec{Q}|^2$ factor and a magnetic form factor, respectively.~\cite{Marshall_1971}

\begin{figure}[htbp]
\begin{center}
\includegraphics[width=0.95\linewidth, keepaspectratio]{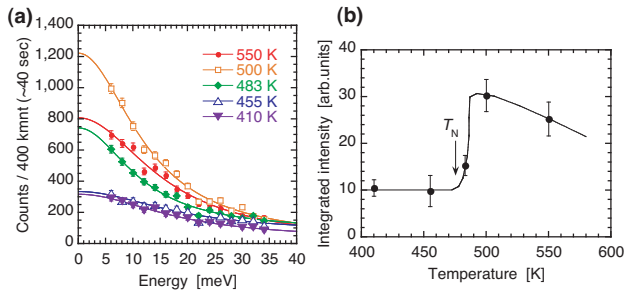}
\end{center}
\caption{\label{fig:others} (Color online) (a) Energy spectra measured at 110 at several temperatures. (b) Temperature dependence of energy-integrated intensity of data in (a). The curves are guides for eye. }
\end{figure}

In the energy spectra measured at 110, as shown in Fig.~\ref{fig:others}(a), the quasielastic tails are extended to over 30 meV.
Figure~\ref{fig:others}(b) shows the temperature dependence of the energy-integrated intensity in Fig.~\ref{fig:others}(a). Here, we see that as the temperature decreases, the intensity grows in the paramagnetic phase, suddenly drops at around $T_{\rm N}$, and finally is sustained at almost the same level in the F-phase.

\section{Discussion}
\subsection{Spatial spin correlations}
\begin{figure}[htbp]
\begin{center}
\includegraphics[width=0.95\linewidth, keepaspectratio]{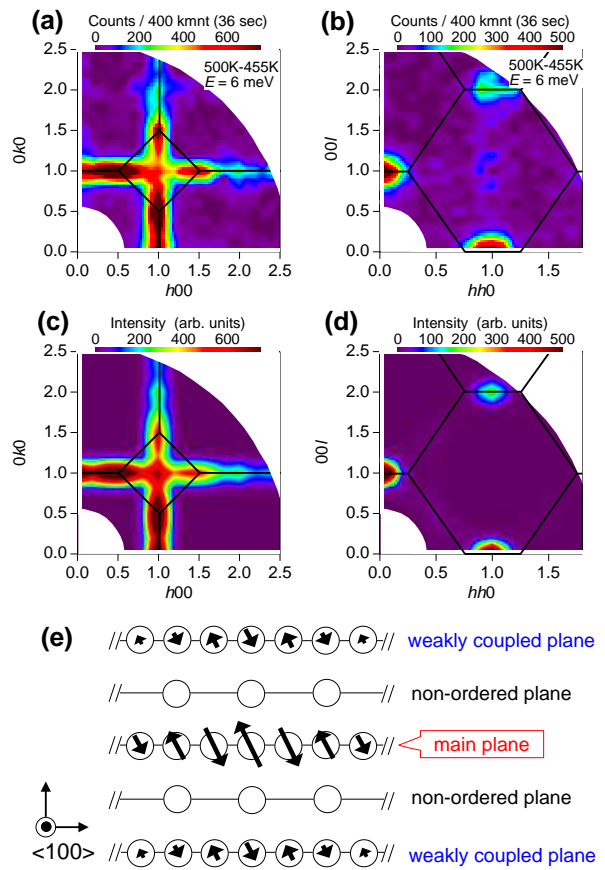}
\end{center}
\caption{\label{fig:model} (Color online) (a)(b) Experimental magnetic scattering intensity at 500 K obtained by subtraction of Fig.~\ref{fig:CEmaps}(e) from Fig.~\ref{fig:CEmaps}(a) and Fig.~\ref{fig:CEmaps}(f) from Fig.~\ref{fig:CEmaps}(b), respectively. (c)(d) Calculated scattering intensity patterns. (e) Corresponding model of spatial spin correlations. The arrows indicate spins that dynamically fluctuate in arbitrary directions. }
\end{figure}

The present experiments revealed that the spatial spin correlations in the paramagnetic phase are roughly described by 2D correlations based on the F-phase magnetic structure. Thus, we searched for a clearer model of spatial correlations that reproduces the experimental intensity patterns.
First, from the experimental scattering intensity at 500 K, we extracted the intensity components that drastically increase above $T_{\rm N}$. That is, by subtracting the data at 455 K (Figs.~\ref{fig:CEmaps}(e) and \ref{fig:CEmaps}(f)) from the data at 500 K (Figs.~\ref{fig:CEmaps}(a) and \ref{fig:CEmaps}(b)), we obtained Figs.~\ref{fig:model}(a) and \ref{fig:model}(b). In the 455 K data shown in Fig.~\ref{fig:CEmaps}(e), the peaky $\vec{q}_{\rm F}$ spots were masked and instead interpolated from their surrounding points. 
Next, we assumed a model based on F-phase magnetic structure, as shown in Fig.~\ref{fig:model}(e). This model consists of three antiferromagnetic planes, where the main plane is sandwiched between the other two vertical planes with a relatively small magnitude of correlated spin components. The intra-layer correlations spatially decay in a Gaussian form, and the spins dynamically fluctuate in arbitrary directions in keeping with the relative correlations.
Then, by adjusting the relative spin magnitude and the intra-layer correlation length, and by using Watson--Freeman's magnetic form factor of Mn,~\cite{Watson_1961} the best-fit intensity patterns were obtained, as shown in Figs.~\ref{fig:model}(c) and \ref{fig:model}(d). They are in a good agreement with Figs.~\ref{fig:model}(a) and \ref{fig:model}(b), respectively.
The magnitude in the two weakly coupled planes was estimated to be 0.1 times that in the main plane, and the intra-layer correlation length was estimated to be six times the first-neighbor Mn--Mn distance in full width at half maximum.

Thus, the temperature dependence of the spatial spin correlations is understood as follows: the inter-layer correlations grow with decreasing temperature, and the same type of static long-range correlations (the F-phase magnetic structure) finally emerges with the tiny tetragonal contraction at $T_{\rm N}$.
Further, suppose there are magnetic interactions only between the ordered sites, as defined in Fig.~\ref{fig:str}(b), the magnetic interactions constituting the spatial correlations can be evaluated. The dominant intra-layer correlations indicate that the first-neighbor interaction $J_1$ is strong and antiferromagnetic, and the additional inter-layer correlations indicate that the second-neighbor interaction $J_2$ is weak and antiferromagnetic as well. This evaluation is consistent with another inelastic neutron scattering study of spin waves in the F-phase, which reported antiferromagnetic $J_{1}=-5.9$ meV and $J_{2}=-1.3$ meV.~\cite{Yamaguchi_1992}

In the F-phase, the dynamical 2D correlations still remain, as shown in Figs.~\ref{fig:CEmaps}(e) and \ref{fig:CEmaps}(f). However, it should be noted that only the non-ordered sites are excluded from these correlations, unlike in the past interpretation.~\cite{Ikeda_2003} This is because the correlations need the square lattice plane consisting of first-neighbor Mn atoms, which does not exist in the non-ordered plane, as shown in Figs.~\ref{fig:str}(b) and \ref{fig:model}(e).

\subsection{Relation between partial disorder and conductivity}
The present experiments revealed that the 2D correlated spin sites and the non-ordered spin sites are separated even in the paramagnetic phase. This result can be related with the conductivity as follows.
Since the paramagnetic phase is cubic, the 2D planes are expected to dynamically change their normal directions among the equivalent $\langle100\rangle$ series. With such changes, the non-ordered sites will be also mobile.
Meanwhile, the coexistence of two types of sites implies the coexistence of different electronic states, which could be accompanied by disproportionality in charge density. In fact, some theoretical studies on itinerant frustrated magnets showed that partial disorder often occurs with charge disproportionation.~\cite{Pinettes_1993,Lacroix_1997,Motome_2010,Hayami_2011} For example, in the Hubbard Hamiltonian, it occurs near the boundary between the nonmagnetic and magnetic phases, when the band width $W$ and onsite Coulomb energy $U$ are close to each other.~\cite{Pinettes_1993,Lacroix_1997} Further, in the Anderson Hamiltonian, it originates from Kondo singlets (Kondo screening).~\cite{Lacroix_1997,Motome_2010,Hayami_2011}
Thus, one can arrive at the following model: the mobile non-ordered sites effectively work as charge carriers and enhance conductivity in the paramagnetic phase.


%
\section{Summary}
We studied spin fluctuations in the frustrated conductor Mn$_3$Pt by single-crystal inelastic neutron scattering. The spatial correlations in the paramagnetic phase are naturally understood as being the inter-layer melt of the F-phase magnetic structure (partial disorder). To the best of our knowledge, such spin fluctuations, and not magnetic structure, were directly observed for the first time. We also proposed a relation between frustration and conductivity: frustration generates non-ordered sites, and they are dynamically mobile and enhance conductivity in the paramagnetic phase. However, an open question remains concerning what electron/hole band structure cause the partial disorder and anomalous conductivity.

\acknowledgments
We thank Messrs. K. Nemoto, M. Ohkawara, and T. Asami for providing assistance at the JAEA, Mr. M. Onodera and Dr. K. Iwasa for their support at Tohoku University, Dr. M. Fujita, Dr. T. Ikeda, and Prof. Y. Tsunoda for the fruitful discussions, and Prof. K. Yamada for the encouragement. The neutron experiments at the JAEA were performed under User Programs conducted by ISSP. This study was financially supported by Grants-in-Aid for Young Scientists (B) (22740209), Priority Areas (22014001), Scientific Researches (S) (21224008) and (A) (22244039) from the MEXT of Japan; it was also supported by the Inter-university Cooperative Research Program of the Institute for Materials Research at Tohoku University.

\bibliography{Mn3Pt_3_PRB}

\end{document}